\documentclass[aps,prl,twocolumn,superscriptaddress, showpacs,floatfix]{revtex4-1}
\usepackage{graphicx}
\usepackage{dcolumn}
\usepackage{bm}
\begin{document}
\title{Skyrmion ground state and gyration of skyrmions in magnetic nanodisks without the Dzyaloshinsky-Moriya interaction}

\author{Yingying Dai}
 \thanks{These authors contributed equally to this work.}
\author{Han Wang}
 \thanks{These authors contributed equally to this work.}
\author{Peng Tao}
\author{Teng Yang}
\author{Weijun Ren}
\author{Zhidong Zhang}
\email[E-mail: ]{zdzhang@imr.ac.cn}
\affiliation{Shenyang National Laboratory for Materials Science,
Institute of Metal Research and International Centre for Materials Physics,
Chinese Academy of Sciences, 72 Wenhua Road, Shenyang 110016,
PRC}
\date{\today}

\begin{abstract}
 We show by micromagnetic simulations that spontaneous skyrmion ground state can exist in Co/Ru/Co nanodisks without the Dzyaloshinsky-Moriya interaction (DMI), which can remain stable in the applied magnetic field along +z direction even up to 0.44 T. The guiding center ($R_x$,$R_y$) of skyrmion defined by the moments of the topological density presents a novel gyration with a star-like trajectory in a pulsed magnetic field and a hexagonal trajectory after the field is switched off, which is different from that of vortex or bubble. One of the coupled skyrmions could move without an external magnetic field, but only induced by the motion of the other one due to strong inter-layer magnetostatic interactions. This work sheds light on how novel skyrmions can be discovered in various (not limited to magnetic) systems with competing energies and contributes to the understanding of the dynamical properties of skyrmion.
\end{abstract}
\pacs{
52.55.-s, 
75.78.Cd, 
12.39.Dc, 
75.50.Ss 
}

\maketitle
\section{introduction}
Skyrmions were originally introduced by the British particle physicist Tony Skyrme to describe localized, particle-like configurations in the field of pion particles.\cite{ref1} Since that seminal paper, skyrmions have been developed in many fields, including classical liquids,\cite{ref2} liquid crystals,\cite{ref3} Bose-Einstein condensates,\cite{ref4} quantum Hall magnets,\cite{ref5} and two-dimensional isotropic magnetic materials.\cite{ref30, ref31, ref32} They also have been theoretically predicted to exist in magnets with the DMI.\cite{ref6} Recently, many experiments have proved their existence in helical magnets, such as MnSi\cite{ref7} and ${\rm Fe_{1-x}Co_xSi}$.\cite{ref8, ref9, ref10} However, most of the skyrmions occurring in helimagnets were induced by an external magnetic field and only at low temperatures,\cite{ref8, ref9, ref10} which limits the technological application of skyrmions.\cite{ref27} Efforts have been made to resolve part of these problems. For example, Yu et al.\cite{ref11} have obtained a skyrmion crystal near room-temperature in the helimagnet FeGe with a high helical transition temperature (280 K), but assisted by an external magnetic field. Heinze et al.\cite{ref12} observed a spontaneous atomic-scale magnetic ground-state skyrmion lattice in a monolayer Fe film, but still at a low temperature (about 11 K). Despite all these efforts, spontaneous skyrmion-like magnetic ground states at or above room temperature have not been reported. Skyrmions existing in helimagnets are always associated with the chiral magnetic interactions, the so-called DMI, which favors canted spin configuration.\cite{ref7, ref8, ref9, ref10, ref13, ref14, ref15, ref16, ref12, ref17, ref18} We note that the magnetostatic interaction favors spin canting and divergence-free spin alignment to reduce the total energy,\cite{ref18, ref52, PhysRevB.86.064427} an effect similar to that of the DMI. For example, confined magnetic thin films can exhibit variant topological spin textures such as vortex-like or meron-like states.\cite{ref19,ref20, ref21, ref22} Therefore, spontaneous ground state of skyrmion-like configurations may be obtainable in ordinary magnets of nanostructures without the DMI.

Topologically nontrivial magnetic nanostructures have attracted long-standing attention due to their peculiar spin configurations and dynamic behaviors for promising technological applications, and in particular to the rich physics involved.\cite{ref27, ref43} Extensive theoretical and experimental research has revealed their dynamics depending on their topological charges and spin textures.\cite{ref39, ref40, ref44} The topological structure of skyrmions distinguishes from magnetic vortex and bubble. Therefore, we expect a novel dynamical behavior in skyrmions.

In this work, we report direct observation of a spontaneous ground state of skyrmion-like configuration in Co/Ru/Co nanodisks without the DMI by micromagnetic simulations. We observe a novel gyration of the guiding center ($R_x$,$R_y$) of skyrmions with a star-like trajectory in a pulsed magnetic field and a hexagonal trajectory after the field is switched off, which is different from the pentagram trajectory of the mean position (X,Y) of magnetic bubbles.\cite{ref46, ref47} Moreover, the gyration of skyrmion in the bottom nanolayer without the presence of field can be stimulated by the motion of skyrmion in the top nanolayer. This indirect control of skyrmion motion offers the potential applications in information-signal processing and spin devices.\cite{ref17, ref27, ref49, ref50, ref51}

\begin{figure}[t]
\includegraphics[width=0.60\columnwidth]{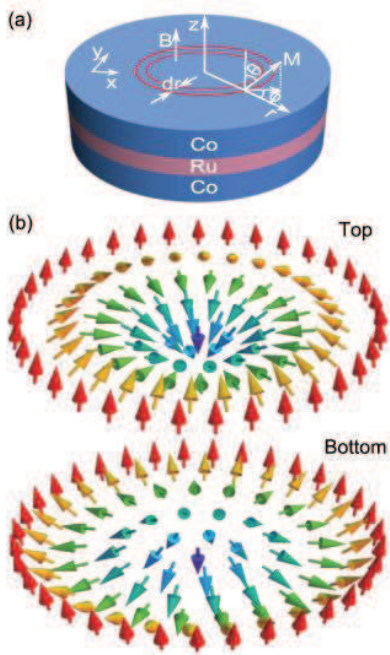}
\caption{(Color online) (a) Sketch of a Co/Ru/Co nanodisk. (b) Micromagnetic simulation result for a Co (20 nm)/Ru (2 nm)/Co (20 nm) nanodisk. Arrows and colors correspond to the directions of the local magnetization and the magnitude of the out-of-plane magnetization component ($M_z$) at every point, respectively. Spin textures in both the top and the bottom nanolayers are skyrmions.
\label{Fig1} }
\end{figure}

\section{micromagnetic methods}
The hexagonal-close-packed (HCP) Co/Ru/Co nanodisks with high Curie temperature and large uniaxial anisotropy are used to study formation and gyrotropic motion of skyrmions without the DMI by the three-dimensional object oriented micromagnetic framework (OOMMF) code.\cite{ref23}
The material parameters of hexagonal-close-packed (HCP) cobalt chosen include saturation magnetization M$_s$=1.4$\times$10$^6$ A/m, exchange stiffness A=3$\times$10$^{-11}$ J/m and uniaxial anisotropy constant K$_u$=5.2$\times$10$^5$ J/m$^3$ with the direction perpendicular to the nanodisk plane. HCP (0001) Co/Ru/Co multilayers can be prepared by electron-beam evaporation, magnetron sputtering deposition or ultrahigh vacuum deposition.\cite{ref35, ref36, ref37} Interfacial coupling coefficients of the adjacent surfaces for different thicknesses of Ru were from Ref. \onlinecite{ref29}. The thickness of Co was tuned from 5 to 25 nm; that of Ru, from 1 to 20 nm. The diameter of a Co/Ru/Co nanodisk in the phase diagram and gyration was chosen to be 200 nm. To study the diameter dependence of the skyrmion radius, the diameter of Co (20 nm)/Ru (2 nm)/Co (20 nm) was from 160 to 210 nm. The cell size was 4$\times$4$\times$1 nm$^3$ for simulating magnetic ground states, which is smaller than the exchange length of cobalt (about 4.94 nm). At phase boundaries, the cell size was reduced to 2$\times$2$\times$1 nm$^3$ to test stability of the obtained states. The dimensionless damping $\alpha$ was chosen to be 0.25 for rapid convergence. Different initial magnetic states [vortex-like (with same or opposite chirality), in-plane-like and out-of-plane-like initial states] were used to get the most stable ground state. As for gyration simulation, the cell size was 2$\times$2$\times$2 nm$^3$ and $\alpha$ was 0.02. A pulsed magnetic field of 10-ns width and 50-mT magnitude along the +x direction was applied to the top or bottom nanolayer.

\section{formation of skyrmions}
\begin{figure*}[t]
\includegraphics[width=0.8\columnwidth]{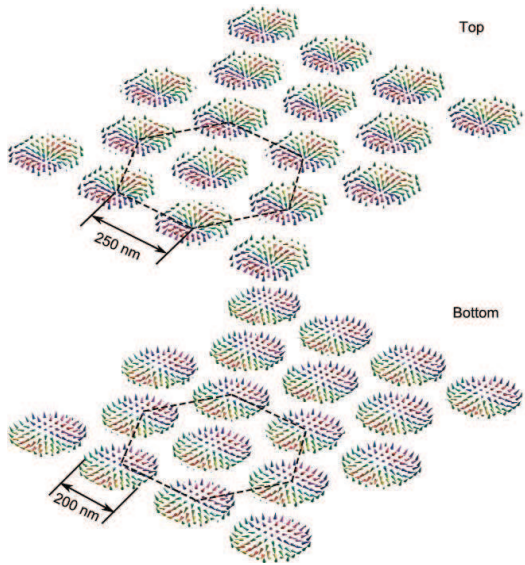}
\caption{(Color online)  Equilibrium magnetic state of a Co (20 nm)/Ru (2 nm)/Co (20 nm) nanodisk array evolving from out-of-plane-like initial state. All the nanodisks have typical skyrmion-like magnetic configurations.
\label{Fig2}}
\end{figure*}

Figure \ref{Fig1}(a) is a sketch of a single Co (20 nm)/Ru (2 nm)/Co (20 nm) nanodisk with a diameter of 200 nm. Figure \ref{Fig1}(b) represents the micromagnetic simulation result of the nanodisk with out-of-plane-like initial state. The equilibrium states of the top and bottom nanolayers are typical skyrmion-like magnetic configurations. The magnetization \textbf{M} is down (along -z axis) in the centers and up (along +z axis) on the boundaries and it rotates gradually from -z axis to +z axis at the intermediate regions of the nanolayers. The magnetic chirality of the top nanolayer is right-handed, while that of the bottom one is left-handed. To elucidate the equilibrium state's nature, we calculate the skyrmion number using the following formula:\cite{ref12, ref46}
\begin{equation}
S = \frac{1}{4\pi}\int\!\!\!\int q dx dy  , \quad q \equiv \frac{1} {2} \epsilon_{\mu \nu} (\partial_\mu \bm{m}\times \partial_\nu \bm{m})\cdot \bm{m}  ,
\label{eq1}
\end{equation}
where $\epsilon_{\mu \nu}$ is the antisymmetric tensor, q is a topological density and $\bm{m}$ is the unit vector of local magnetization. S is found to be approximately -1, showing a signature of skyrmion-like state.
Similar magnetic spin textures have been found in a patterned Co/Ru/Co nanodisk array with the diameter the same as that of the single Co/Ru/Co nanodisk. The distance between centers of two nearest neighboring nanodisks was 250 nm, as shown in Fig. \ref{Fig2}. The result suggests that a stray field between two nearest neighboring nanodisks (250 nm apart) has little influence on the skyrmion spin textures.

\begin{figure}[t]
\includegraphics[width=0.8\columnwidth]{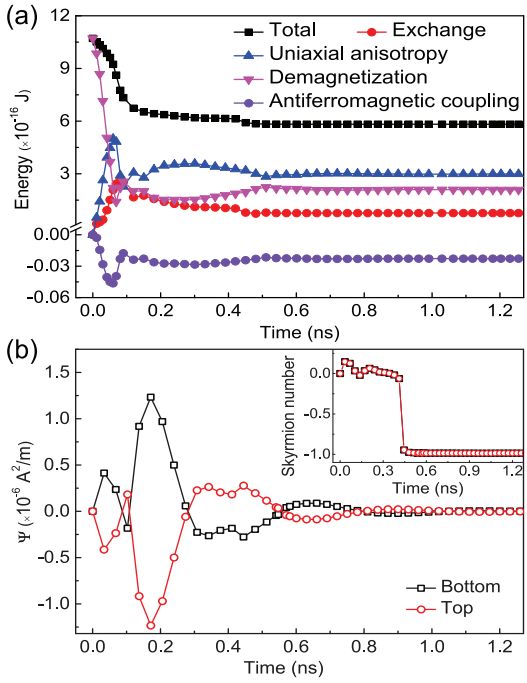}
\caption{(Color online) (a) Time dependence of different energies for a Co (20 nm)/Ru (2 nm)/Co (20 nm) nanodisk with a diameter of 200 nm. (b) Competition among the exchange energy, demagnetization energy and uniaxial anisotropy energy has an effect similar to the DMI (here represented by $\Psi$). Inset: the skyrmion number as a function of time.
\label{Fig3} }
\end{figure}

The formation of magnetic stable states is the consequence of minimizing the Gibbs free energy of magnetic systems. Figure \ref{Fig3}(a) illustrates time dependences of the total energy, exchange energy, uniaxial anisotropy energy, demagnetization energy, and antiferromagnetic coupling energy for the case in Fig. \ref{Fig1}. The exchange energy, uniaxial anisotropy energy, and demagnetization energy are two orders of magnitude higher than the interfacial antiferromagnetic coupling energy and, thus, are vital to the emergence of a skyrmion spin texture. At the beginning, the out-of-plane-like initial state has a very high value of total energy due to the significant demagnetization energy. To lower the total energy, the demagnetization energy decreases rapidly with time, whereas the exchange energy and uniaxial anisotropy energy both increase significantly. A balance is reached and the total energy is almost unchanged after 0.5 ns. Concurrently, the skyrmion number S drops to about -1 for the top and bottom nanolayers, as shown in the inset in Fig. \ref{Fig3}(b), suggesting that a spontaneous topologically stable knot emerges in the magnetization.\cite{ref7}

As elaborated in many articles,\cite{ref7, ref8, ref9, ref10,ref11, ref13, ref14, ref15, ref16, ref12, ref17, ref18} the DMI is crucial to the formation of magnetic skyrmion-like states, which favors canting spins. The DMI is defined as \cite{ref9, ref24, ref25, ref26}
\begin{equation}
H_{DMI} = \int\!\!\!\int D \bm{M} \cdot (\nabla \times \bm{M})dxdy  ,
\label{eq2}
\end{equation}
where \emph{D} is DMI constant. But in this work, skyrmions are spontaneously formed without the DMI. As discussed above,
the competition among the exchange energy, demagnetization energy, and uniaxial
anisotropy energy plays a significant role in the emergence of skyrmions. To quantify the competition effect, we define a quantity to mimic the DMI:
\begin{equation}
\Psi = \int\!\!\!\int \bm{M} \cdot (\nabla \times \bm{M})dxdy  .
\label{eq3}
\end{equation}
From Eq. (\ref{eq3}), we calculate $\Psi$ as a function of time on both the top and the bottom Co nanolayers [see Fig. \ref{Fig3}(b)]. Notably, when energies compete drastically with each other before 0.5 ns in Fig. \ref{Fig3}(a), The $\Psi$ for both Co nanolayers significantly changes. Then all energies reach equilibrium and $\Psi$ remains relatively unchanged after 0.8 ns. This indicates that the competition among the three energies can produce an effect similar to that of the DMI to form skyrmions.

\begin{figure}[b]
\includegraphics[width=0.7\columnwidth]{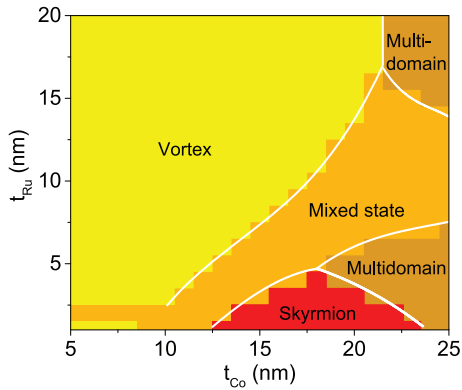}
\caption{(Color online) Phase diagram of spin textures derived from micromagnetic simulations. The spin textures as functions of the thickness of Co and of Ru illustrate four regions: the vortex-like state, skyrmion-like state, multidomain state, and mixed state. Phase boundaries are marked by white lines.
\label{Fig4} }
\end{figure}

Figure \ref{Fig4} shows the phase diagram of the magnetic ground state of a Co/Ru/Co nanodisk as a function of the thicknesses of ruthenium ($t_{Ru}$) and cobalt ($t_{Co}$) derived from micromagnetic simulations. $t_{Ru}$ is changed from 1 to 20 nm, and $t_{Co}$ from 5 to 25 nm, with a fixed diameter of 200 nm. The phase diagram shows four regions: the vortex-like state, skyrmion-like state, multidomain state and mixed state (more than one stable or metastable state).
Stable skyrmions can exist only in a small fraction of the phase diagram with $t_{Co}$ appropriately chosen and $t_{Ru}$ smaller than 4 nm. This is obviously a delicate balance between different energies, especially the interlayer magnetostatic interaction. If $t_{Co}$ goes beyond 25 nm, regardless of $t_{Ru}$, the multidomain state appears to minimize the demagnetization energy. If the separation layer of Ru gets very thick where the interlayer magnetostatic interaction between two Co nanolayers becomes diluted, the intralayer demagnetization factor starts to play a more essential role, constraining the local magnetization in plane to form the vortex state. Though we have been emphasizing the importance of interlayer magnetostatic interaction by controlling the $t_{Ru}$, it is still not sufficient to reduce just the $t_{Ru}$ for the skyrmion state. Moreover, the interfacial antiferromagnetic or ferromagnetic coupling becomes more effective once $t_{Co}$ is smaller than 8 nm.


\section{stability of skyrmion}

\begin{figure}[t]
\includegraphics[width=0.8\columnwidth]{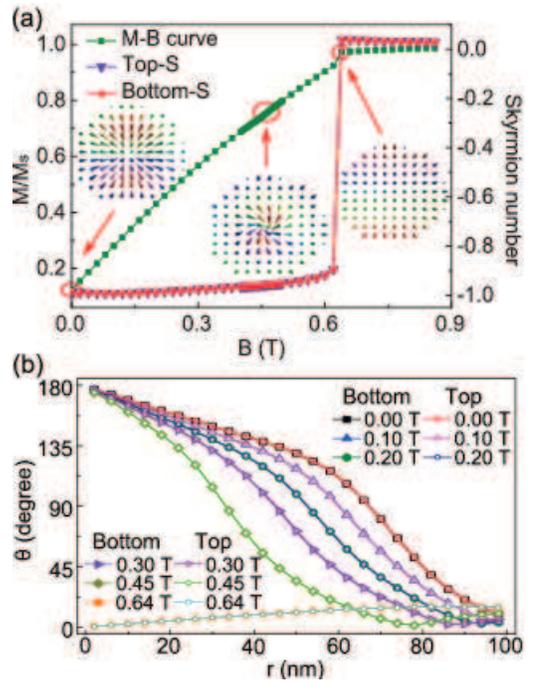}
\caption{(Color online) (a) Normalized magnetization curve (M-B curve) of the Co (20 nm)/Ru (2 nm)/Co (20 nm) nanodisk and the corresponding skyrmion number of the top ("Top-S") and the bottom ("Bottom-S") nanolayers as a function of the field. (b) Averaged angle between the local magnetization and the +z axis in the range of (r, r+dr) under different magnetic fields.
\label{Fig5} }
\end{figure}

Figure \ref{Fig5} represents the stability of skyrmions under an external magnetic field along +z axis. The normalized magnetization curve of the nanodisk and the corresponding skyrmion number as a function of the field are shown in Fig. \ref{Fig5}(a). The insets in Fig. \ref{Fig5}(a) illustrate the profiles of magnetic stable states of the top nanolayer at different fields. The magnetization increases almost linearly with the magnetic field until saturation at 0.64 T, where skyrmions are suppressed completely. Concurrently, skyrmion numbers of the top and bottom nanolayers change slightly, from approximately -1 to -0.9, and rise sharply, up to about 0, when the field is up to 0.64 T. To investigate the spatial distribution of magnetization, we show the \textit{r} dependence of $\theta$ in Figs. \ref{Fig5}(b). $\theta$ is defined as the averaged angle between the local magnetization and +z axis in the range of (r, r+dr) as described in Fig. \ref{Fig1}(a). The behavior of $\theta$ at zero magnetic field is typically skyrmion-like and rules out the possible existence of cylindrical bubble domains.\cite{ref27, ref53} Besides, the distribution of magnetization does not match the Belavin-Polyakov (BP) \cite{ref30} solution as demonstrated in Fig. \ref{Fig9}. With an increasing magnetic field, $\theta$ values in the center and at the edge of the nanodisk align rigidly till the critical saturation field (0.64 T) is reached. $\theta$ at the edge remains always tilted from a 0 value, possibly due to the stray field at the edge, which results in the absence of a small fraction of the order parameter sphere when the skyrmion is mapped from real space to an order parameter sphere as illustrated in Fig. \ref{Fig8}. In contrast to the rigidity of magnetization in the center and at the edge, $\theta$ in the intermediate region decreases gradually with the field, which has a variation tendency similar to that reported in Ref. \onlinecite{ref28}. Figure \ref{Fig5} indicates that skyrmions remain in a similar configuration under the external field even up to 0.44 T, which is much higher than the reported one\cite{ref7, ref8, ref9, ref11}. Obviously, the skyrmion obtained in the nanodisk is quite robust.

\begin{figure*}[t]
\includegraphics[width=1.0\columnwidth]{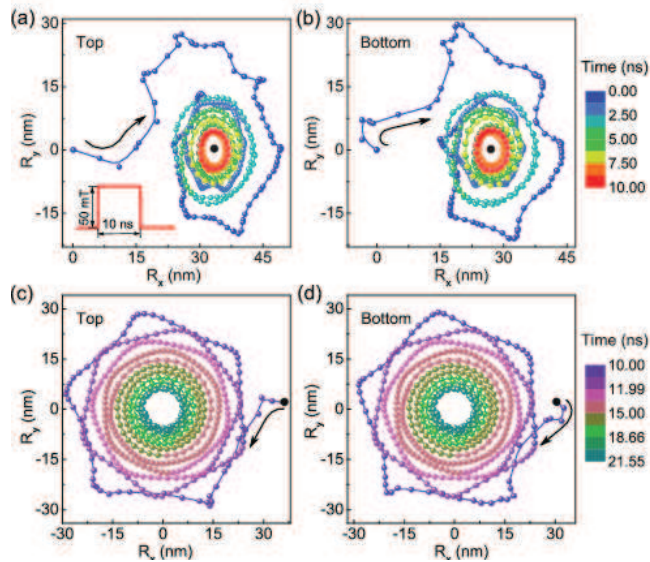}
\caption{(Color online) (a, b) Trajectories of the guiding centers of skyrmions in both the top and the bottom nanolayers when a pulsed magnetic field is applied to the top nanolayer. (c), d) Gyrotropic motion of the guiding centers of skyrmions after the applied magnetic field is switched off. Colors are used to indicate time-dependent positions of ($R_x$,$R_y$). The filled black circle represents the new equilibrium position under the magnetic field.
\label{Fig6} }
\end{figure*}

\section{dynamics of skyrmions}

\begin{figure*}[t]
\includegraphics[width=1.\columnwidth]{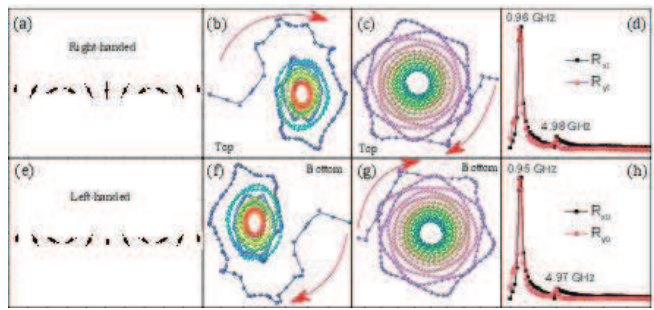}
\caption{(Color online) Influence of skyrmion chiralities on gyration of the guiding center of skyrmions. (a, e) schematics of skyrmions with right-handed and left-handed chiralities. (b, f) motion of opposite chiral skyrmions when a pulsed magnetic field is applied along the +x direction to the top or bottom nanolayer, respectively. (c, g) gyrotropic motion of skyrmions after the applied field is switched off. (d),(h) Corresponding FFT spectra of the hexagonal trajectories. (R$_{xt}$, R$_{yt}$) and (R$_{xb}$, R$_{yb}$) are the positions of the guiding center of skyrmions in the top and the bottom nanolayer, respectively.
\label{Fig7} }
\end{figure*}

\begin{figure}[b]
\includegraphics[width=0.8\columnwidth]{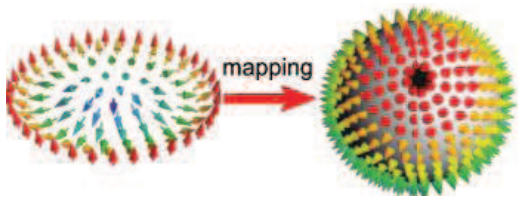}
\caption{(Color online) Mapping from the skyrmion configuration to the unit sphere.
\label{Fig8} }
\end{figure}

We also investigated the gyrotropic motion of the guiding center of skyrmions. The guiding center ($R_x$,$R_y$) is essential for the dynamics of a skyrmion, which is defined by the moments of the topological density:\cite{ref41, ref46}
\begin{equation}
R_x = \frac{\int\!\!\!\int x q dx dy} {\int\!\!\!\int q dx dy} , \quad R_y = \frac{\int\!\!\!\int y q dx dy} {\int\!\!\!\int q dx dy},
\label{eq4}
\end{equation}
where q is the topological density defined in Eq. (\ref{eq1}).
First, a pulsed magnetic field of 10-ns width and 50-mT magnitude is applied along the +x direction on the top nanolayer, as demonstrated in Fig. \ref{Fig6}(a). The guiding center gyrates towards its new equlibrium position along the field, which is different from the motion of the vortex.\cite{ref33, ref34} The trajectory of ($R_x$,$R_y$) of the top nanolayer is initially like a star (before 2.4 ns), then an elliptical orbit, and, finally, is damped around its new equilibrium position (33.5 nm, 0 nm), represented by the filled black circle.
Stimulated by the gyrotropic motion of the top skyrmion through strong interlayer magnetostatic interaction, the skyrmions in the bottom nanolayer without the presence of a field almost synchronously gyrate, as shown in Fig. \ref{Fig6}(b). This indirect control of skyrmion motion may be used in information-signal processing and spin devices.\cite{ref17, ref27, ref49, ref50, ref51}
Once the field is turned off, ($R_x$,$R_y$) in both nanolayers begin to gyrate back to the origin (0,0), as shown in Figs. \ref{Fig6}(c) and \ref{Fig6}(d). Initally, the trajectories of the two guiding centers of skyrmions are hexagons (before 15.24 ns), which is similar to hypocycloid involute. The period of this hexagon-like motion is approximately T=1 ns (i.e., frequency \emph{f}=1 GHz). The average velocity in the first period is about 200 m/s. The corresponding FFT spectra of the novel trajectories illustrate two eigenfrequencies of 0.96 and 4.98 GHz with a ratio of about 1:5 as illustrated in Fig. \ref{Fig7}.
These unique trajectories of ($R_x$,$R_y$) have never been observed in other topological magnetic systems (vortices or bubbles), which usually have a circular or elliptical orbit. Note that a pentagram trajectory resulting from the deformation of circular domain wall in magnetic bubble was reported previously,\cite{ref46, ref47} which is about the mean position (X,Y) defined by the moments of the magnetization. However, hexagon-like trajectories in this work are due to the distribution variance of the topological density of the whole system and about the guiding centers ($R_x$,$R_y$) defined by the moments of the topological density. Moreover, the motion of skyrmion can not be correctly described by the mean position (X,Y). Therefore, our hexagon-like trajectories of ($R_x$,$R_y$) in skyrmion is sharply different from the pentagram trajectory of (X,Y) in magnetic bubble. These may be understood by their differences in topological charges and spin textures as well as the strong magnetostatic interaction of the double-skyrmion system (detailed analysis will be discussed in future work). After 14.24 ns, gyration orbits change to circular ones.

The influence of skyrmion chirality on the guiding center of skyrmion gyration is illustrated in Fig. \ref{Fig7}. Figures \ref{Fig7}(a) and \ref{Fig7}(e) illustrate sketches of right-handed and left-handed magnetization. The skyrmion of top and bottom nanolayers are right-handed and left-handed, respectively. The motion of opposite chiral skyrmions when a pulsed magnetic field along +x direction is applied on the top or bottom nanolayer are shown in Figs. \ref{Fig7}(b) and \ref{Fig7}(f). The guiding center of skyrmion gyrates towards new equilibrium position along the field applied on the top nanolayer, while against the field when applying the field on the bottom nanolayer. Figures \ref{Fig7}(c) and \ref{Fig7}(g) are the corresponding gyrations after the applied field is switched off. The trajectories are similar. Both of the trajectories are hexagons before 15.24 ns and the corresponding fast Fourier transform (FFT) spectra are demonstrated in Figs. \ref{Fig7}(d) and \ref{Fig7}(h). Both of the hexagonal trajectories have two eigenfrequencies with approximately the same values, about 0.96 and 4.98 GHz. The ratio of the two frequencies is about 1:5.

\section{discussion}

Figure \ref{Fig8} shows the mapping to the unit sphere of an order parameter space from the skyrmion configuration of a Co (20 nm)/Ru (2 nm)/Co (20 nm) nanodisk with a diameter of 200 nm. A small fraction of a region is missing from the unit sphere. This missing region corresponds to about 0.5\% of the sphere area. Nevertheless, the mapped region can still give rise to a skyrmion number (about -0.995) close to -1.0 for a fully mapped skyrmion sphere.

\begin{figure}[b]
\includegraphics[width=0.8\columnwidth]{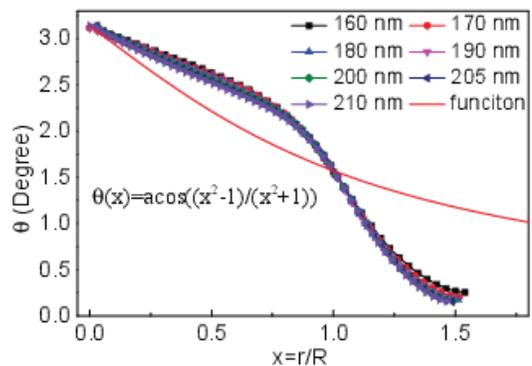}
\caption{(Color online) Variation of $\theta$ as a function of x (x=r/R) for different diameters of Co (20 nm)/Ru (2 nm)/Co (20 nm) nanodisks. The solid (red) line is the BP solution of the nonlinear SO(3) $\sigma$-model.
\label{Fig9} }
\end{figure}

$\theta$ as a function of x (x=r/R) is illustrated in Fig. \ref{Fig9}, where R is the skyrmion radius in which $\theta$ goes to $\pi$/2. Diameters of Co/Ru/Co nanodisks are from 160 to 210 nm. All curves from different nanodisk diameters collapse to one.  The skyrmion radius is found to be from 51 to 70 nm, corresponding to a diameter of from 160 to 210 nm. The solid (red) line is the BP solution for the nonlinear SO(3) $\sigma$ model. Skyrmions obtained in the Co/Ru/Co nanodisks do not match the BP solution. In these nanodisks, the competition among exchange energy, uniaxial anisotropy energy, and demagnetization energy results in the formation of a skyrmion spin configuration instead of the single exchange energy considered in the BP solution.

\section{conclusion}
In conclusion, we have revealed the existence of a spontaneous magnetic skyrmion ground state in Co/Ru/Co nanodisks without chiral DMI. The competition of the exchange energy, demagnetization energy, and uniaxial anisotropy energy acts similarly to the DMI and thus leads to the formation of an intriguing skyrmion texture. This nontrivial topological spin texture is very stable even against an external magnetic field up to 0.44 T, higher than the highest reported value, about 0.3 T. We observe a novel gyration of the guiding centers ($R_x$,$R_y$) with a star-like trajectory in a magnetic field and a hexagonal trajectory after the field is switched off. We also observe synchronous gyrotropic motions of two skyrmions in two Co nanolayers (one with a field, the other one without). These unique gyrotropic motions of the skyrmion are distinguished from the magnetic vortex or bubble and contribute to the understanding of the dynamical properties of skyrmions. Our study suggests a new method to the generation of skyrmion textures in various confined magnetic systems and paves the new way to the design of spintronic and magnetic storage devices.

\begin{acknowledgments}
The work was supported by the National Basic Research Program (No. 2010CB934603) of China, Ministry of Science and Technology China and the National Natural Science Foundation of China under Grant No. 50831006.
\end{acknowledgments}


\begin{thebibliography}{50}%
\makeatletter
\providecommand \@ifxundefined [1]{%
 \@ifx{#1\undefined}
}%
\providecommand \@ifnum [1]{%
 \ifnum #1\expandafter \@firstoftwo
 \else \expandafter \@secondoftwo
 \fi
}%
\providecommand \@ifx [1]{%
 \ifx #1\expandafter \@firstoftwo
 \else \expandafter \@secondoftwo
 \fi
}%
\providecommand \natexlab [1]{#1}%
\providecommand \enquote  [1]{``#1''}%
\providecommand \bibnamefont  [1]{#1}%
\providecommand \bibfnamefont [1]{#1}%
\providecommand \citenamefont [1]{#1}%
\providecommand \href@noop [0]{\@secondoftwo}%
\providecommand \href [0]{\begingroup \@sanitize@url \@href}%
\providecommand \@href[1]{\@@startlink{#1}\@@href}%
\providecommand \@@href[1]{\endgroup#1\@@endlink}%
\providecommand \@sanitize@url [0]{\catcode `\\12\catcode `\$12\catcode
  `\&12\catcode `\#12\catcode `\^12\catcode `\_12\catcode `\%12\relax}%
\providecommand \@@startlink[1]{}%
\providecommand \@@endlink[0]{}%
\providecommand \url  [0]{\begingroup\@sanitize@url \@url }%
\providecommand \@url [1]{\endgroup\@href {#1}{\urlprefix }}%
\providecommand \urlprefix  [0]{URL }%
\providecommand \Eprint [0]{\href }%
\providecommand \doibase [0]{http://dx.doi.org/}%
\providecommand \selectlanguage [0]{\@gobble}%
\providecommand \bibinfo  [0]{\@secondoftwo}%
\providecommand \bibfield  [0]{\@secondoftwo}%
\providecommand \translation [1]{[#1]}%
\providecommand \BibitemOpen [0]{}%
\providecommand \bibitemStop [0]{}%
\providecommand \bibitemNoStop [0]{.\EOS\space}%
\providecommand \EOS [0]{\spacefactor3000\relax}%
\providecommand \BibitemShut  [1]{\csname bibitem#1\endcsname}%
\let\auto@bib@innerbib\@empty
\bibitem [{\citenamefont {Skyrme}(1962)}]{ref1}%
  \BibitemOpen
  \bibfield  {author} {\bibinfo {author} {\bibfnamefont {T.~H.~R.}\
  \bibnamefont {Skyrme}},\ }\href@noop {} {\bibfield  {journal} {\bibinfo
  {journal} {Nucl. Phys.}\ }\textbf {\bibinfo {volume} {31}},\ \bibinfo {pages}
  {556} (\bibinfo {year} {1962})}\BibitemShut {NoStop}%
\bibitem [{\citenamefont {Cross}\ and\ \citenamefont {Hohenberg}(1993)}]{ref2}%
  \BibitemOpen
  \bibfield  {author} {\bibinfo {author} {\bibfnamefont {M.~C.}\ \bibnamefont
  {Cross}}\ and\ \bibinfo {author} {\bibfnamefont {P.~C.}\ \bibnamefont
  {Hohenberg}},\ }\href {\doibase 10.1103/RevModPhys.65.851} {\bibfield
  {journal} {\bibinfo  {journal} {Rev. Mod. Phys.}\ }\textbf {\bibinfo {volume}
  {65}},\ \bibinfo {pages} {851} (\bibinfo {year} {1993})}\BibitemShut
  {NoStop}%
\bibitem [{\citenamefont {Wright}\ and\ \citenamefont {Mermin}(1989)}]{ref3}%
  \BibitemOpen
  \bibfield  {author} {\bibinfo {author} {\bibfnamefont {D.~C.}\ \bibnamefont
  {Wright}}\ and\ \bibinfo {author} {\bibfnamefont {N.~D.}\ \bibnamefont
  {Mermin}},\ }\href {\doibase 10.1103/RevModPhys.61.385} {\bibfield  {journal}
  {\bibinfo  {journal} {Rev. Mod. Phys.}\ }\textbf {\bibinfo {volume} {61}},\
  \bibinfo {pages} {385} (\bibinfo {year} {1989})}\BibitemShut {NoStop}%
\bibitem [{\citenamefont {Al~Khawaja}\ and\ \citenamefont
  {Stoof}(2001)}]{ref4}%
  \BibitemOpen
  \bibfield  {author} {\bibinfo {author} {\bibfnamefont {U.}~\bibnamefont
  {Al~Khawaja}}\ and\ \bibinfo {author} {\bibfnamefont {H.}~\bibnamefont
  {Stoof}},\ }\href {\doibase 10.1038/35082010} {\bibfield  {journal} {\bibinfo
   {journal} {Nature}\ }\textbf {\bibinfo {volume} {411}},\ \bibinfo {pages}
  {918} (\bibinfo {year} {2001})}\BibitemShut {NoStop}%
\bibitem [{\citenamefont {Sondhi}\ \emph {et~al.}(1993)\citenamefont {Sondhi},
  \citenamefont {Karlhede}, \citenamefont {Kivelson},\ and\ \citenamefont
  {Rezayi}}]{ref5}%
  \BibitemOpen
  \bibfield  {author} {\bibinfo {author} {\bibfnamefont {S.~L.}\ \bibnamefont
  {Sondhi}}, \bibinfo {author} {\bibfnamefont {A.}~\bibnamefont {Karlhede}},
  \bibinfo {author} {\bibfnamefont {S.~A.}\ \bibnamefont {Kivelson}}, \ and\
  \bibinfo {author} {\bibfnamefont {E.~H.}\ \bibnamefont {Rezayi}},\ }\href
  {\doibase 10.1103/PhysRevB.47.16419} {\bibfield  {journal} {\bibinfo
  {journal} {Phys. Rev. B}\ }\textbf {\bibinfo {volume} {47}},\ \bibinfo
  {pages} {16419} (\bibinfo {year} {1993})}\BibitemShut {NoStop}%
\bibitem [{\citenamefont {Belavin}\ and\ \citenamefont
  {Polyakov}(1975)}]{ref30}%
  \BibitemOpen
  \bibfield  {author} {\bibinfo {author} {\bibfnamefont {A.~A.}\ \bibnamefont
  {Belavin}}\ and\ \bibinfo {author} {\bibfnamefont {A.~M.}\ \bibnamefont
  {Polyakov}},\ }\href@noop {} {\bibfield  {journal} {\bibinfo  {journal} {JETP
  Lett.}\ }\textbf {\bibinfo {volume} {{22}}},\ \bibinfo {pages} {{245}}
  (\bibinfo {year} {{1975}})}\BibitemShut {NoStop}%
\bibitem [{\citenamefont {Zaspel}\ \emph {et~al.}(1995)\citenamefont {Zaspel},
  \citenamefont {Grigereit},\ and\ \citenamefont {Drumheller}}]{ref31}%
  \BibitemOpen
  \bibfield  {author} {\bibinfo {author} {\bibfnamefont {C.~E.}\ \bibnamefont
  {Zaspel}}, \bibinfo {author} {\bibfnamefont {T.~E.}\ \bibnamefont
  {Grigereit}}, \ and\ \bibinfo {author} {\bibfnamefont {J.~E.}\ \bibnamefont
  {Drumheller}},\ }\href {\doibase 10.1103/PhysRevLett.74.4539} {\bibfield
  {journal} {\bibinfo  {journal} {Phys. Rev. Lett.}\ }\textbf {\bibinfo
  {volume} {74}},\ \bibinfo {pages} {4539} (\bibinfo {year}
  {1995})}\BibitemShut {NoStop}%
\bibitem [{\citenamefont {Waldner}(1986)}]{ref32}%
  \BibitemOpen
  \bibfield  {author} {\bibinfo {author} {\bibfnamefont {F.}~\bibnamefont
  {Waldner}},\ }\href {\doibase 10.1016/0304-8853(86)90293-3} {\bibfield
  {journal} {\bibinfo  {journal} {J. Magn. Magn. Mater.}\ }\textbf {\bibinfo
  {volume} {54-57}},\ \bibinfo {pages} {873} (\bibinfo {year}
  {1986})}\BibitemShut {NoStop}%
\bibitem [{\citenamefont {Bogdanov}\ and\ \citenamefont
  {Yablonskii}(1989)}]{ref6}%
  \BibitemOpen
  \bibfield  {author} {\bibinfo {author} {\bibfnamefont {A.~N.}\ \bibnamefont
  {Bogdanov}}\ and\ \bibinfo {author} {\bibfnamefont {D.~A.}\ \bibnamefont
  {Yablonskii}},\ }\href@noop {} {\bibfield  {journal} {\bibinfo  {journal}
  {Sov. Phys. JETP}\ }\textbf {\bibinfo {volume} {68}},\ \bibinfo {pages} {101}
  (\bibinfo {year} {1989})}\BibitemShut {NoStop}%
\bibitem [{\citenamefont {M$\ddot{u}$hlbauer}\ \emph
  {et~al.}(2009)\citenamefont {M$\ddot{u}$hlbauer}, \citenamefont {Binz},
  \citenamefont {Jonietz}, \citenamefont {Pfleiderer}, \citenamefont {Rosch},
  \citenamefont {Neubauer}, \citenamefont {Georgii},\ and\ \citenamefont
  {B$\ddot{o}$ni}}]{ref7}%
  \BibitemOpen
  \bibfield  {author} {\bibinfo {author} {\bibfnamefont {S.}~\bibnamefont
  {M$\ddot{u}$hlbauer}}, \bibinfo {author} {\bibfnamefont {B.}~\bibnamefont
  {Binz}}, \bibinfo {author} {\bibfnamefont {F.}~\bibnamefont {Jonietz}},
  \bibinfo {author} {\bibfnamefont {C.}~\bibnamefont {Pfleiderer}}, \bibinfo
  {author} {\bibfnamefont {A.}~\bibnamefont {Rosch}}, \bibinfo {author}
  {\bibfnamefont {A.}~\bibnamefont {Neubauer}}, \bibinfo {author}
  {\bibfnamefont {R.}~\bibnamefont {Georgii}}, \ and\ \bibinfo {author}
  {\bibfnamefont {P.}~\bibnamefont {B$\ddot{o}$ni}},\ }\href@noop {} {\bibfield
   {journal} {\bibinfo  {journal} {Science}\ }\textbf {\bibinfo {volume}
  {323}},\ \bibinfo {pages} {915} (\bibinfo {year} {2009})}\BibitemShut
  {NoStop}%
\bibitem [{\citenamefont {M$\ddot{u}$nzer}\ \emph {et~al.}(2010)\citenamefont
  {M$\ddot{u}$nzer}, \citenamefont {Neubauer}, \citenamefont {Adams},
  \citenamefont {M$\ddot{u}$hlbauer}, \citenamefont {Franz}, \citenamefont
  {Jonietz}, \citenamefont {Georgii}, \citenamefont {B$\ddot{o}$ni},
  \citenamefont {Pedersen}, \citenamefont {Schmidt}, \citenamefont {Rosch},\
  and\ \citenamefont {Pfleiderer}}]{ref8}%
  \BibitemOpen
  \bibfield  {author} {\bibinfo {author} {\bibfnamefont {W.}~\bibnamefont
  {M$\ddot{u}$nzer}}, \bibinfo {author} {\bibfnamefont {A.}~\bibnamefont
  {Neubauer}}, \bibinfo {author} {\bibfnamefont {T.}~\bibnamefont {Adams}},
  \bibinfo {author} {\bibfnamefont {S.}~\bibnamefont {M$\ddot{u}$hlbauer}},
  \bibinfo {author} {\bibfnamefont {C.}~\bibnamefont {Franz}}, \bibinfo
  {author} {\bibfnamefont {F.}~\bibnamefont {Jonietz}}, \bibinfo {author}
  {\bibfnamefont {R.}~\bibnamefont {Georgii}}, \bibinfo {author} {\bibfnamefont
  {P.}~\bibnamefont {B$\ddot{o}$ni}}, \bibinfo {author} {\bibfnamefont
  {B.}~\bibnamefont {Pedersen}}, \bibinfo {author} {\bibfnamefont
  {M.}~\bibnamefont {Schmidt}}, \bibinfo {author} {\bibfnamefont
  {A.}~\bibnamefont {Rosch}}, \ and\ \bibinfo {author} {\bibfnamefont
  {C.}~\bibnamefont {Pfleiderer}},\ }\href {\doibase
  10.1103/PhysRevB.81.041203} {\bibfield  {journal} {\bibinfo  {journal} {Phys.
  Rev. B}\ }\textbf {\bibinfo {volume} {81}},\ \bibinfo {pages} {041203}
  (\bibinfo {year} {2010})}\BibitemShut {NoStop}%
\bibitem [{\citenamefont {Yu}\ \emph {et~al.}(2010)\citenamefont {Yu},
  \citenamefont {Onose}, \citenamefont {Kanazawa}, \citenamefont {Park},
  \citenamefont {Han}, \citenamefont {Matsui}, \citenamefont {Nagaosa},\ and\
  \citenamefont {Tokura}}]{ref9}%
  \BibitemOpen
  \bibfield  {author} {\bibinfo {author} {\bibfnamefont {X.~Z.}\ \bibnamefont
  {Yu}}, \bibinfo {author} {\bibfnamefont {Y.}~\bibnamefont {Onose}}, \bibinfo
  {author} {\bibfnamefont {N.}~\bibnamefont {Kanazawa}}, \bibinfo {author}
  {\bibfnamefont {J.~H.}\ \bibnamefont {Park}}, \bibinfo {author}
  {\bibfnamefont {J.~H.}\ \bibnamefont {Han}}, \bibinfo {author} {\bibfnamefont
  {Y.}~\bibnamefont {Matsui}}, \bibinfo {author} {\bibfnamefont
  {N.}~\bibnamefont {Nagaosa}}, \ and\ \bibinfo {author} {\bibfnamefont
  {Y.}~\bibnamefont {Tokura}},\ }\href {\doibase 10.1038/nature09124}
  {\bibfield  {journal} {\bibinfo  {journal} {Nature}\ }\textbf {\bibinfo
  {volume} {465}},\ \bibinfo {pages} {901} (\bibinfo {year}
  {2010})}\BibitemShut {NoStop}%
\bibitem [{\citenamefont {Tonomura}\ \emph {et~al.}(2012)\citenamefont
  {Tonomura}, \citenamefont {Yu}, \citenamefont {Yanagisawa}, \citenamefont
  {Matsuda}, \citenamefont {Onose}, \citenamefont {Kanazawa}, \citenamefont
  {Park},\ and\ \citenamefont {Tokura}}]{ref10}%
  \BibitemOpen
  \bibfield  {author} {\bibinfo {author} {\bibfnamefont {A.}~\bibnamefont
  {Tonomura}}, \bibinfo {author} {\bibfnamefont {X.}~\bibnamefont {Yu}},
  \bibinfo {author} {\bibfnamefont {K.}~\bibnamefont {Yanagisawa}}, \bibinfo
  {author} {\bibfnamefont {T.}~\bibnamefont {Matsuda}}, \bibinfo {author}
  {\bibfnamefont {Y.}~\bibnamefont {Onose}}, \bibinfo {author} {\bibfnamefont
  {N.}~\bibnamefont {Kanazawa}}, \bibinfo {author} {\bibfnamefont {H.~S.}\
  \bibnamefont {Park}}, \ and\ \bibinfo {author} {\bibfnamefont
  {Y.}~\bibnamefont {Tokura}},\ }\href {\doibase 10.1021/nl300073m} {\bibfield
  {journal} {\bibinfo  {journal} {Nano Lett.}\ }\textbf {\bibinfo {volume}
  {12}},\ \bibinfo {pages} {1673} (\bibinfo {year} {2012})}\BibitemShut
  {NoStop}%
\bibitem [{\citenamefont {Kiselev}\ \emph
  {et~al.}(2011{\natexlab{a}})\citenamefont {Kiselev}, \citenamefont
  {Bogdanov}, \citenamefont {Sch$\ddot{a}$fer},\ and\ \citenamefont
  {R$\ddot{o}$$\ss$ler}}]{ref27}%
  \BibitemOpen
  \bibfield  {author} {\bibinfo {author} {\bibfnamefont {N.~S.}\ \bibnamefont
  {Kiselev}}, \bibinfo {author} {\bibfnamefont {A.~N.}\ \bibnamefont
  {Bogdanov}}, \bibinfo {author} {\bibfnamefont {R.}~\bibnamefont
  {Sch$\ddot{a}$fer}}, \ and\ \bibinfo {author} {\bibfnamefont {U.~K.}\
  \bibnamefont {R$\ddot{o}$$\ss$ler}},\ }\href@noop {} {\bibfield  {journal}
  {\bibinfo  {journal} {J. Phys. D: Appl. Phys.}\ }\textbf {\bibinfo {volume}
  {44}},\ \bibinfo {pages} {392001} (\bibinfo {year}
  {2011}{\natexlab{a}})}\BibitemShut {NoStop}%
\bibitem [{\citenamefont {Yu}\ \emph {et~al.}(2011)\citenamefont {Yu},
  \citenamefont {Kanazawa}, \citenamefont {Onose}, \citenamefont {Kimoto},
  \citenamefont {Zhang}, \citenamefont {Ishiwata}, \citenamefont {Matsui},\
  and\ \citenamefont {Tokura}}]{ref11}%
  \BibitemOpen
  \bibfield  {author} {\bibinfo {author} {\bibfnamefont {X.~Z.}\ \bibnamefont
  {Yu}}, \bibinfo {author} {\bibfnamefont {N.}~\bibnamefont {Kanazawa}},
  \bibinfo {author} {\bibfnamefont {Y.}~\bibnamefont {Onose}}, \bibinfo
  {author} {\bibfnamefont {K.}~\bibnamefont {Kimoto}}, \bibinfo {author}
  {\bibfnamefont {W.~Z.}\ \bibnamefont {Zhang}}, \bibinfo {author}
  {\bibfnamefont {S.}~\bibnamefont {Ishiwata}}, \bibinfo {author}
  {\bibfnamefont {Y.}~\bibnamefont {Matsui}}, \ and\ \bibinfo {author}
  {\bibfnamefont {Y.}~\bibnamefont {Tokura}},\ }\href {\doibase
  10.1038/nmat2916} {\bibfield  {journal} {\bibinfo  {journal} {Nat. Mater.}\
  }\textbf {\bibinfo {volume} {10}},\ \bibinfo {pages} {106 } (\bibinfo {year}
  {2011})}\BibitemShut {NoStop}%
\bibitem [{\citenamefont {Heinze}\ \emph {et~al.}(2011)\citenamefont {Heinze},
  \citenamefont {von Bergmann}, \citenamefont {Menzel}, \citenamefont {Brede},
  \citenamefont {Kubetzka}, \citenamefont {Wiesendanger}, \citenamefont
  {Bihlmayer},\ and\ \citenamefont {Bl$\ddot{u}$gel}}]{ref12}%
  \BibitemOpen
  \bibfield  {author} {\bibinfo {author} {\bibfnamefont {S.}~\bibnamefont
  {Heinze}}, \bibinfo {author} {\bibfnamefont {K.}~\bibnamefont {von
  Bergmann}}, \bibinfo {author} {\bibfnamefont {M.}~\bibnamefont {Menzel}},
  \bibinfo {author} {\bibfnamefont {J.}~\bibnamefont {Brede}}, \bibinfo
  {author} {\bibfnamefont {A.}~\bibnamefont {Kubetzka}}, \bibinfo {author}
  {\bibfnamefont {R.}~\bibnamefont {Wiesendanger}}, \bibinfo {author}
  {\bibfnamefont {G.}~\bibnamefont {Bihlmayer}}, \ and\ \bibinfo {author}
  {\bibfnamefont {S.}~\bibnamefont {Bl$\ddot{u}$gel}},\ }\href {\doibase
  10.1038/nphys2045} {\bibfield  {journal} {\bibinfo  {journal} {Nat. Phys.}\
  }\textbf {\bibinfo {volume} {7}},\ \bibinfo {pages} {713 } (\bibinfo {year}
  {2011})}\BibitemShut {NoStop}%
\bibitem [{\citenamefont {Adams}\ \emph {et~al.}(2010)\citenamefont {Adams},
  \citenamefont {M$\ddot{u}$hlbauer}, \citenamefont {Neubauer}, \citenamefont
  {M$\ddot{u}$nzer}, \citenamefont {Jonietz}, \citenamefont {Georgii},
  \citenamefont {Pedersen}, \citenamefont {B$\ddot{o}$ni}, \citenamefont
  {Rosch},\ and\ \citenamefont {Pfleiderer}}]{ref13}%
  \BibitemOpen
  \bibfield  {author} {\bibinfo {author} {\bibfnamefont {T.}~\bibnamefont
  {Adams}}, \bibinfo {author} {\bibfnamefont {S.}~\bibnamefont
  {M$\ddot{u}$hlbauer}}, \bibinfo {author} {\bibfnamefont {A.}~\bibnamefont
  {Neubauer}}, \bibinfo {author} {\bibfnamefont {W.}~\bibnamefont
  {M$\ddot{u}$nzer}}, \bibinfo {author} {\bibfnamefont {F.}~\bibnamefont
  {Jonietz}}, \bibinfo {author} {\bibfnamefont {R.}~\bibnamefont {Georgii}},
  \bibinfo {author} {\bibfnamefont {B.}~\bibnamefont {Pedersen}}, \bibinfo
  {author} {\bibfnamefont {P.}~\bibnamefont {B$\ddot{o}$ni}}, \bibinfo {author}
  {\bibfnamefont {A.}~\bibnamefont {Rosch}}, \ and\ \bibinfo {author}
  {\bibfnamefont {C.}~\bibnamefont {Pfleiderer}},\ }\href@noop {} {\bibfield
  {journal} {\bibinfo  {journal} {J. Phys.: Conf. Ser.}\ }\textbf {\bibinfo
  {volume} {200}},\ \bibinfo {pages} {032001} (\bibinfo {year}
  {2010})}\BibitemShut {NoStop}%
\bibitem [{\citenamefont {Adams}\ \emph {et~al.}(2011)\citenamefont {Adams},
  \citenamefont {M$\ddot{u}$hlbauer}, \citenamefont {Pfleiderer}, \citenamefont
  {Jonietz}, \citenamefont {Bauer}, \citenamefont {Neubauer}, \citenamefont
  {Georgii}, \citenamefont {B$\ddot{o}$ni}, \citenamefont {Keiderling},
  \citenamefont {Everschor}, \citenamefont {Garst},\ and\ \citenamefont
  {Rosch}}]{ref14}%
  \BibitemOpen
  \bibfield  {author} {\bibinfo {author} {\bibfnamefont {T.}~\bibnamefont
  {Adams}}, \bibinfo {author} {\bibfnamefont {S.}~\bibnamefont
  {M$\ddot{u}$hlbauer}}, \bibinfo {author} {\bibfnamefont {C.}~\bibnamefont
  {Pfleiderer}}, \bibinfo {author} {\bibfnamefont {F.}~\bibnamefont {Jonietz}},
  \bibinfo {author} {\bibfnamefont {A.}~\bibnamefont {Bauer}}, \bibinfo
  {author} {\bibfnamefont {A.}~\bibnamefont {Neubauer}}, \bibinfo {author}
  {\bibfnamefont {R.}~\bibnamefont {Georgii}}, \bibinfo {author} {\bibfnamefont
  {P.}~\bibnamefont {B$\ddot{o}$ni}}, \bibinfo {author} {\bibfnamefont
  {U.}~\bibnamefont {Keiderling}}, \bibinfo {author} {\bibfnamefont
  {K.}~\bibnamefont {Everschor}}, \bibinfo {author} {\bibfnamefont
  {M.}~\bibnamefont {Garst}}, \ and\ \bibinfo {author} {\bibfnamefont
  {A.}~\bibnamefont {Rosch}},\ }\href {\doibase 10.1103/PhysRevLett.107.217206}
  {\bibfield  {journal} {\bibinfo  {journal} {Phys. Rev. Lett.}\ }\textbf
  {\bibinfo {volume} {107}},\ \bibinfo {pages} {217206} (\bibinfo {year}
  {2011})}\BibitemShut {NoStop}%
\bibitem [{\citenamefont {Bogdanov}\ and\ \citenamefont
  {R$\ddot{o}$${\ss}$ler}(2001)}]{ref15}%
  \BibitemOpen
  \bibfield  {author} {\bibinfo {author} {\bibfnamefont {A.~N.}\ \bibnamefont
  {Bogdanov}}\ and\ \bibinfo {author} {\bibfnamefont {U.~K.}\ \bibnamefont
  {R$\ddot{o}$${\ss}$ler}},\ }\href {\doibase 10.1103/PhysRevLett.87.037203}
  {\bibfield  {journal} {\bibinfo  {journal} {Phys. Rev. Lett.}\ }\textbf
  {\bibinfo {volume} {87}},\ \bibinfo {pages} {037203} (\bibinfo {year}
  {2001})}\BibitemShut {NoStop}%
\bibitem [{\citenamefont {Bogdanov}\ and\ \citenamefont
  {Hubert}(1994)}]{ref16}%
  \BibitemOpen
  \bibfield  {author} {\bibinfo {author} {\bibfnamefont {A.}~\bibnamefont
  {Bogdanov}}\ and\ \bibinfo {author} {\bibfnamefont {A.}~\bibnamefont
  {Hubert}},\ }\href {\doibase 10.1016/0304-8853(94)90046-9} {\bibfield
  {journal} {\bibinfo  {journal} {J. Magn. Magn. Mater.}\ }\textbf {\bibinfo
  {volume} {138}},\ \bibinfo {pages} {255} (\bibinfo {year}
  {1994})}\BibitemShut {NoStop}%
\bibitem [{\citenamefont {Pfleiderer}\ and\ \citenamefont
  {Rosch}(2010)}]{ref17}%
  \BibitemOpen
  \bibfield  {author} {\bibinfo {author} {\bibfnamefont {C.}~\bibnamefont
  {Pfleiderer}}\ and\ \bibinfo {author} {\bibfnamefont {A.}~\bibnamefont
  {Rosch}},\ }\href {\doibase 10.1038/465880a} {\bibfield  {journal} {\bibinfo
  {journal} {Nature}\ }\textbf {\bibinfo {volume} {465}},\ \bibinfo {pages}
  {880 } (\bibinfo {year} {2010})}\BibitemShut {NoStop}%
\bibitem [{\citenamefont {Chung}\ \emph {et~al.}(2010)\citenamefont {Chung},
  \citenamefont {McMichael}, \citenamefont {Pierce},\ and\ \citenamefont
  {Unguris}}]{ref18}%
  \BibitemOpen
  \bibfield  {author} {\bibinfo {author} {\bibfnamefont {S.-H.}\ \bibnamefont
  {Chung}}, \bibinfo {author} {\bibfnamefont {R.~D.}\ \bibnamefont
  {McMichael}}, \bibinfo {author} {\bibfnamefont {D.~T.}\ \bibnamefont
  {Pierce}}, \ and\ \bibinfo {author} {\bibfnamefont {J.}~\bibnamefont
  {Unguris}},\ }\href {\doibase 10.1103/PhysRevB.81.024410} {\bibfield
  {journal} {\bibinfo  {journal} {Phys. Rev. B}\ }\textbf {\bibinfo {volume}
  {81}},\ \bibinfo {pages} {024410} (\bibinfo {year} {2010})}\BibitemShut
  {NoStop}%
\bibitem [{\citenamefont {Ezawa}(2010)}]{ref52}%
  \BibitemOpen
  \bibfield  {author} {\bibinfo {author} {\bibfnamefont {M.}~\bibnamefont
  {Ezawa}},\ }\href {\doibase 10.1103/PhysRevLett.105.197202} {\bibfield
  {journal} {\bibinfo  {journal} {Phys. Rev. Lett.}\ }\textbf {\bibinfo
  {volume} {105}},\ \bibinfo {pages} {197202} (\bibinfo {year}
  {2010})}\BibitemShut {NoStop}%
\bibitem [{\citenamefont {Johnson}\ \emph {et~al.}(2012)\citenamefont
  {Johnson}, \citenamefont {Gangopadhyay}, \citenamefont {Kalyanaraman},\ and\
  \citenamefont {Nussinov}}]{PhysRevB.86.064427}%
  \BibitemOpen
  \bibfield  {author} {\bibinfo {author} {\bibfnamefont {P.}~\bibnamefont
  {Johnson}}, \bibinfo {author} {\bibfnamefont {A.~K.}\ \bibnamefont
  {Gangopadhyay}}, \bibinfo {author} {\bibfnamefont {R.}~\bibnamefont
  {Kalyanaraman}}, \ and\ \bibinfo {author} {\bibfnamefont {Z.}~\bibnamefont
  {Nussinov}},\ }\href {\doibase 10.1103/PhysRevB.86.064427} {\bibfield
  {journal} {\bibinfo  {journal} {Phys. Rev. B}\ }\textbf {\bibinfo {volume}
  {86}},\ \bibinfo {pages} {064427} (\bibinfo {year} {2012})}\BibitemShut
  {NoStop}%
\bibitem [{\citenamefont {Cowburn}\ \emph {et~al.}(1999)\citenamefont
  {Cowburn}, \citenamefont {Koltsov}, \citenamefont {Adeyeye}, \citenamefont
  {Welland},\ and\ \citenamefont {Tricker}}]{ref19}%
  \BibitemOpen
  \bibfield  {author} {\bibinfo {author} {\bibfnamefont {R.~P.}\ \bibnamefont
  {Cowburn}}, \bibinfo {author} {\bibfnamefont {D.~K.}\ \bibnamefont
  {Koltsov}}, \bibinfo {author} {\bibfnamefont {A.~O.}\ \bibnamefont
  {Adeyeye}}, \bibinfo {author} {\bibfnamefont {M.~E.}\ \bibnamefont
  {Welland}}, \ and\ \bibinfo {author} {\bibfnamefont {D.~M.}\ \bibnamefont
  {Tricker}},\ }\href {\doibase 10.1103/PhysRevLett.83.1042} {\bibfield
  {journal} {\bibinfo  {journal} {Phys. Rev. Lett.}\ }\textbf {\bibinfo
  {volume} {83}},\ \bibinfo {pages} {1042} (\bibinfo {year}
  {1999})}\BibitemShut {NoStop}%
\bibitem [{\citenamefont {Phatak}\ \emph {et~al.}(2012)\citenamefont {Phatak},
  \citenamefont {Petford-Long},\ and\ \citenamefont {Heinonen}}]{ref20}%
  \BibitemOpen
  \bibfield  {author} {\bibinfo {author} {\bibfnamefont {C.}~\bibnamefont
  {Phatak}}, \bibinfo {author} {\bibfnamefont {A.~K.}\ \bibnamefont
  {Petford-Long}}, \ and\ \bibinfo {author} {\bibfnamefont {O.}~\bibnamefont
  {Heinonen}},\ }\href {\doibase 10.1103/PhysRevLett.108.067205} {\bibfield
  {journal} {\bibinfo  {journal} {Phys. Rev. Lett.}\ }\textbf {\bibinfo
  {volume} {108}},\ \bibinfo {pages} {067205} (\bibinfo {year}
  {2012})}\BibitemShut {NoStop}%
\bibitem [{\citenamefont {Shinjo}\ \emph {et~al.}(2000)\citenamefont {Shinjo},
  \citenamefont {Okuno}, \citenamefont {Hassdorf}, \citenamefont {Shigeto},\
  and\ \citenamefont {Ono}}]{ref21}%
  \BibitemOpen
  \bibfield  {author} {\bibinfo {author} {\bibfnamefont {T.}~\bibnamefont
  {Shinjo}}, \bibinfo {author} {\bibfnamefont {T.}~\bibnamefont {Okuno}},
  \bibinfo {author} {\bibfnamefont {R.}~\bibnamefont {Hassdorf}}, \bibinfo
  {author} {\bibfnamefont {K.}~\bibnamefont {Shigeto}}, \ and\ \bibinfo
  {author} {\bibfnamefont {T.}~\bibnamefont {Ono}},\ }\href {\doibase
  10.1126/science.289.5481.930} {\bibfield  {journal} {\bibinfo  {journal}
  {Science}\ }\textbf {\bibinfo {volume} {289}},\ \bibinfo {pages} {930}
  (\bibinfo {year} {2000})}\BibitemShut {NoStop}%
\bibitem [{\citenamefont {Van~Waeyenberge}\ \emph {et~al.}(2006)\citenamefont
  {Van~Waeyenberge}, \citenamefont {Puzic}, \citenamefont {Stoll},
  \citenamefont {Chou}, \citenamefont {Tyliszczak}, \citenamefont {Hertel},
  \citenamefont {F$\ddot{a}$hnle}, \citenamefont {Br$\ddot{u}$ckl},
  \citenamefont {Rott}, \citenamefont {Reiss}, \citenamefont {Neudecker},
  \citenamefont {Weiss}, \citenamefont {Back},\ and\ \citenamefont
  {Schutz}}]{ref22}%
  \BibitemOpen
  \bibfield  {author} {\bibinfo {author} {\bibfnamefont {B.}~\bibnamefont
  {Van~Waeyenberge}}, \bibinfo {author} {\bibfnamefont {A.}~\bibnamefont
  {Puzic}}, \bibinfo {author} {\bibfnamefont {H.}~\bibnamefont {Stoll}},
  \bibinfo {author} {\bibfnamefont {K.~W.}\ \bibnamefont {Chou}}, \bibinfo
  {author} {\bibfnamefont {T.}~\bibnamefont {Tyliszczak}}, \bibinfo {author}
  {\bibfnamefont {R.}~\bibnamefont {Hertel}}, \bibinfo {author} {\bibfnamefont
  {M.}~\bibnamefont {F$\ddot{a}$hnle}}, \bibinfo {author} {\bibfnamefont
  {H.}~\bibnamefont {Br$\ddot{u}$ckl}}, \bibinfo {author} {\bibfnamefont
  {K.}~\bibnamefont {Rott}}, \bibinfo {author} {\bibfnamefont {G.}~\bibnamefont
  {Reiss}}, \bibinfo {author} {\bibfnamefont {I.}~\bibnamefont {Neudecker}},
  \bibinfo {author} {\bibfnamefont {D.}~\bibnamefont {Weiss}}, \bibinfo
  {author} {\bibfnamefont {C.~H.}\ \bibnamefont {Back}}, \ and\ \bibinfo
  {author} {\bibfnamefont {G.}~\bibnamefont {Schutz}},\ }\href {\doibase
  10.1038/nature05240} {\bibfield  {journal} {\bibinfo  {journal} {Nature}\
  }\textbf {\bibinfo {volume} {444}},\ \bibinfo {pages} {461 } (\bibinfo {year}
  {2006})}\BibitemShut {NoStop}%
\bibitem [{\citenamefont {Mart\'in}\ \emph {et~al.}(2003)\citenamefont
  {Mart\'in}, \citenamefont {Nogu\'es}, \citenamefont {Liu}, \citenamefont
  {Vicent},\ and\ \citenamefont {Schuller}}]{ref43}%
  \BibitemOpen
  \bibfield  {author} {\bibinfo {author} {\bibfnamefont {J.}~\bibnamefont
  {Mart\'in}}, \bibinfo {author} {\bibfnamefont {J.}~\bibnamefont {Nogu\'es}},
  \bibinfo {author} {\bibfnamefont {K.}~\bibnamefont {Liu}}, \bibinfo {author}
  {\bibfnamefont {J.}~\bibnamefont {Vicent}}, \ and\ \bibinfo {author}
  {\bibfnamefont {I.~K.}\ \bibnamefont {Schuller}},\ }\href {\doibase
  10.1016/S0304-8853(02)00898-3} {\bibfield  {journal} {\bibinfo  {journal} {J.
  Magn. Magn. Mater.}\ }\textbf {\bibinfo {volume} {256}},\ \bibinfo {pages}
  {449 } (\bibinfo {year} {2003})}\BibitemShut {NoStop}%
\bibitem [{\citenamefont {Dussaux}\ \emph {et~al.}(2010)\citenamefont
  {Dussaux}, \citenamefont {Georges}, \citenamefont {Grollier}, \citenamefont
  {Cros}, \citenamefont {Khvalkovskiy}, \citenamefont {Fukushima},
  \citenamefont {Konoto}, \citenamefont {Kubota}, \citenamefont {Yakushiji},
  \citenamefont {Yuasa}, \citenamefont {Zvezdin}, \citenamefont {Ando},\ and\
  \citenamefont {Fert}}]{ref39}%
  \BibitemOpen
  \bibfield  {author} {\bibinfo {author} {\bibfnamefont {A.}~\bibnamefont
  {Dussaux}}, \bibinfo {author} {\bibfnamefont {B.}~\bibnamefont {Georges}},
  \bibinfo {author} {\bibfnamefont {J.}~\bibnamefont {Grollier}}, \bibinfo
  {author} {\bibfnamefont {V.}~\bibnamefont {Cros}}, \bibinfo {author}
  {\bibfnamefont {A.~V.}\ \bibnamefont {Khvalkovskiy}}, \bibinfo {author}
  {\bibfnamefont {A.}~\bibnamefont {Fukushima}}, \bibinfo {author}
  {\bibfnamefont {M.}~\bibnamefont {Konoto}}, \bibinfo {author} {\bibfnamefont
  {H.}~\bibnamefont {Kubota}}, \bibinfo {author} {\bibfnamefont
  {K.}~\bibnamefont {Yakushiji}}, \bibinfo {author} {\bibfnamefont
  {S.}~\bibnamefont {Yuasa}}, \bibinfo {author} {\bibfnamefont {K.~A.}\
  \bibnamefont {Zvezdin}}, \bibinfo {author} {\bibfnamefont {K.}~\bibnamefont
  {Ando}}, \ and\ \bibinfo {author} {\bibfnamefont {A.}~\bibnamefont {Fert}},\
  }\href@noop {} {\bibfield  {journal} {\bibinfo  {journal} {{Nat. Commun.}}\
  }\textbf {\bibinfo {volume} {{1}}},\ \bibinfo {pages} {{8}} (\bibinfo {year}
  {{2010}})}\BibitemShut {NoStop}%
\bibitem [{\citenamefont {Pribiag}\ \emph {et~al.}(2007)\citenamefont
  {Pribiag}, \citenamefont {Krivorotov}, \citenamefont {Fuchs}, \citenamefont
  {Braganca}, \citenamefont {Ozatay}, \citenamefont {Sankey}, \citenamefont
  {Ralph},\ and\ \citenamefont {Buhrman}}]{ref40}%
  \BibitemOpen
  \bibfield  {author} {\bibinfo {author} {\bibfnamefont {V.~S.}\ \bibnamefont
  {Pribiag}}, \bibinfo {author} {\bibfnamefont {I.~N.}\ \bibnamefont
  {Krivorotov}}, \bibinfo {author} {\bibfnamefont {G.~D.}\ \bibnamefont
  {Fuchs}}, \bibinfo {author} {\bibfnamefont {P.~M.}\ \bibnamefont {Braganca}},
  \bibinfo {author} {\bibfnamefont {O.}~\bibnamefont {Ozatay}}, \bibinfo
  {author} {\bibfnamefont {J.~C.}\ \bibnamefont {Sankey}}, \bibinfo {author}
  {\bibfnamefont {D.~C.}\ \bibnamefont {Ralph}}, \ and\ \bibinfo {author}
  {\bibfnamefont {R.~A.}\ \bibnamefont {Buhrman}},\ }\href@noop {} {\bibfield
  {journal} {\bibinfo  {journal} {Nat. Phys.}\ }\textbf {\bibinfo {volume}
  {3}},\ \bibinfo {pages} {498} (\bibinfo {year} {2007})}\BibitemShut {NoStop}%
\bibitem [{\citenamefont {Kasai}\ \emph {et~al.}(2006)\citenamefont {Kasai},
  \citenamefont {Nakatani}, \citenamefont {Kobayashi}, \citenamefont {Kohno},\
  and\ \citenamefont {Ono}}]{ref44}%
  \BibitemOpen
  \bibfield  {author} {\bibinfo {author} {\bibfnamefont {S.}~\bibnamefont
  {Kasai}}, \bibinfo {author} {\bibfnamefont {Y.}~\bibnamefont {Nakatani}},
  \bibinfo {author} {\bibfnamefont {K.}~\bibnamefont {Kobayashi}}, \bibinfo
  {author} {\bibfnamefont {H.}~\bibnamefont {Kohno}}, \ and\ \bibinfo {author}
  {\bibfnamefont {T.}~\bibnamefont {Ono}},\ }\href {\doibase
  10.1103/PhysRevLett.97.107204} {\bibfield  {journal} {\bibinfo  {journal}
  {Phys. Rev. Lett.}\ }\textbf {\bibinfo {volume} {97}},\ \bibinfo {pages}
  {107204} (\bibinfo {year} {2006})}\BibitemShut {NoStop}%
\bibitem [{\citenamefont {Moutafis}\ \emph {et~al.}(2009)\citenamefont
  {Moutafis}, \citenamefont {Komineas},\ and\ \citenamefont {Bland}}]{ref46}%
  \BibitemOpen
  \bibfield  {author} {\bibinfo {author} {\bibfnamefont {C.}~\bibnamefont
  {Moutafis}}, \bibinfo {author} {\bibfnamefont {S.}~\bibnamefont {Komineas}},
  \ and\ \bibinfo {author} {\bibfnamefont {J.~A.~C.}\ \bibnamefont {Bland}},\
  }\href {\doibase 10.1103/PhysRevB.79.224429} {\bibfield  {journal} {\bibinfo
  {journal} {Phys. Rev. B}\ }\textbf {\bibinfo {volume} {79}},\ \bibinfo
  {pages} {224429} (\bibinfo {year} {2009})}\BibitemShut {NoStop}%
\bibitem [{\citenamefont {Makhfudz}\ \emph {et~al.}(2012)\citenamefont
  {Makhfudz}, \citenamefont {Kr\"uger},\ and\ \citenamefont
  {Tchernyshyov}}]{ref47}%
  \BibitemOpen
  \bibfield  {author} {\bibinfo {author} {\bibfnamefont {I.}~\bibnamefont
  {Makhfudz}}, \bibinfo {author} {\bibfnamefont {B.}~\bibnamefont {Kr\"uger}},
  \ and\ \bibinfo {author} {\bibfnamefont {O.}~\bibnamefont {Tchernyshyov}},\
  }\href {\doibase 10.1103/PhysRevLett.109.217201} {\bibfield  {journal}
  {\bibinfo  {journal} {Phys. Rev. Lett.}\ }\textbf {\bibinfo {volume} {109}},\
  \bibinfo {pages} {217201} (\bibinfo {year} {2012})}\BibitemShut {NoStop}%
\bibitem [{\citenamefont {Jung}\ \emph {et~al.}(2011)\citenamefont {Jung},
  \citenamefont {Lee}, \citenamefont {Jeong}, \citenamefont {Choi},
  \citenamefont {Yu}, \citenamefont {Han}, \citenamefont {Vogel}, \citenamefont
  {Bocklage}, \citenamefont {Meier}, \citenamefont {Im}, \citenamefont
  {Fischer},\ and\ \citenamefont {Kim}}]{ref49}%
  \BibitemOpen
  \bibfield  {author} {\bibinfo {author} {\bibfnamefont {H.}~\bibnamefont
  {Jung}}, \bibinfo {author} {\bibfnamefont {K.-S.}\ \bibnamefont {Lee}},
  \bibinfo {author} {\bibfnamefont {D.-E.}\ \bibnamefont {Jeong}}, \bibinfo
  {author} {\bibfnamefont {Y.-S.}\ \bibnamefont {Choi}}, \bibinfo {author}
  {\bibfnamefont {Y.-S.}\ \bibnamefont {Yu}}, \bibinfo {author} {\bibfnamefont
  {D.-S.}\ \bibnamefont {Han}}, \bibinfo {author} {\bibfnamefont
  {A.}~\bibnamefont {Vogel}}, \bibinfo {author} {\bibfnamefont
  {L.}~\bibnamefont {Bocklage}}, \bibinfo {author} {\bibfnamefont
  {G.}~\bibnamefont {Meier}}, \bibinfo {author} {\bibfnamefont {M.-Y.}\
  \bibnamefont {Im}}, \bibinfo {author} {\bibfnamefont {P.}~\bibnamefont
  {Fischer}}, \ and\ \bibinfo {author} {\bibfnamefont {S.-K.}\ \bibnamefont
  {Kim}},\ }\href {\doibase {10.1038/srep00059}} {\bibfield  {journal}
  {\bibinfo  {journal} {{Scientific Reports}}\ }\textbf {\bibinfo {volume}
  {{1}}},\ \bibinfo {pages} {{59}} (\bibinfo {year} {{2011}})}\BibitemShut
  {NoStop}%
\bibitem [{\citenamefont {Wieser}\ \emph {et~al.}(2011)\citenamefont {Wieser},
  \citenamefont {Vedmedenko},\ and\ \citenamefont {Wiesendanger}}]{ref50}%
  \BibitemOpen
  \bibfield  {author} {\bibinfo {author} {\bibfnamefont {R.}~\bibnamefont
  {Wieser}}, \bibinfo {author} {\bibfnamefont {E.~Y.}\ \bibnamefont
  {Vedmedenko}}, \ and\ \bibinfo {author} {\bibfnamefont {R.}~\bibnamefont
  {Wiesendanger}},\ }\href {\doibase 10.1103/PhysRevLett.106.067204} {\bibfield
   {journal} {\bibinfo  {journal} {Phys. Rev. Lett.}\ }\textbf {\bibinfo
  {volume} {106}},\ \bibinfo {pages} {067204} (\bibinfo {year}
  {2011})}\BibitemShut {NoStop}%
\bibitem [{\citenamefont {O'Brien}\ \emph {et~al.}(2012)\citenamefont
  {O'Brien}, \citenamefont {Lewis}, \citenamefont {Fern\'andez-Pacheco},
  \citenamefont {Petit}, \citenamefont {Cowburn}, \citenamefont {Sampaio},\
  and\ \citenamefont {Read}}]{ref51}%
  \BibitemOpen
  \bibfield  {author} {\bibinfo {author} {\bibfnamefont {L.}~\bibnamefont
  {O'Brien}}, \bibinfo {author} {\bibfnamefont {E.~R.}\ \bibnamefont {Lewis}},
  \bibinfo {author} {\bibfnamefont {A.}~\bibnamefont {Fern\'andez-Pacheco}},
  \bibinfo {author} {\bibfnamefont {D.}~\bibnamefont {Petit}}, \bibinfo
  {author} {\bibfnamefont {R.~P.}\ \bibnamefont {Cowburn}}, \bibinfo {author}
  {\bibfnamefont {J.}~\bibnamefont {Sampaio}}, \ and\ \bibinfo {author}
  {\bibfnamefont {D.~E.}\ \bibnamefont {Read}},\ }\href {\doibase
  10.1103/PhysRevLett.108.187202} {\bibfield  {journal} {\bibinfo  {journal}
  {Phys. Rev. Lett.}\ }\textbf {\bibinfo {volume} {108}},\ \bibinfo {pages}
  {187202} (\bibinfo {year} {2012})}\BibitemShut {NoStop}%
\bibitem [{\citenamefont {Donahue}\ and\ \citenamefont {Porter}(1999)}]{ref23}%
  \BibitemOpen
  \bibfield  {author} {\bibinfo {author} {\bibfnamefont {M.~J.}\ \bibnamefont
  {Donahue}}\ and\ \bibinfo {author} {\bibfnamefont {D.~G.}\ \bibnamefont
  {Porter}},\ }\href {http://math.nist.gov/oommf} {\emph {\bibinfo {title}
  {NISTIR 6376}}}\ (\bibinfo {year} {1999})\BibitemShut {NoStop}%
\bibitem [{\citenamefont {Ounadjela}\ \emph {et~al.}(1992)\citenamefont
  {Ounadjela}, \citenamefont {Muller}, \citenamefont {Dinia}, \citenamefont
  {Arbaoui}, \citenamefont {Panissod},\ and\ \citenamefont {Suran}}]{ref35}%
  \BibitemOpen
  \bibfield  {author} {\bibinfo {author} {\bibfnamefont {K.}~\bibnamefont
  {Ounadjela}}, \bibinfo {author} {\bibfnamefont {D.}~\bibnamefont {Muller}},
  \bibinfo {author} {\bibfnamefont {A.}~\bibnamefont {Dinia}}, \bibinfo
  {author} {\bibfnamefont {A.}~\bibnamefont {Arbaoui}}, \bibinfo {author}
  {\bibfnamefont {P.}~\bibnamefont {Panissod}}, \ and\ \bibinfo {author}
  {\bibfnamefont {G.}~\bibnamefont {Suran}},\ }\href {\doibase
  10.1103/PhysRevB.45.7768} {\bibfield  {journal} {\bibinfo  {journal} {Phys.
  Rev. B}\ }\textbf {\bibinfo {volume} {45}},\ \bibinfo {pages} {7768}
  (\bibinfo {year} {1992})}\BibitemShut {NoStop}%
\bibitem [{\citenamefont {Alayo}\ \emph {et~al.}(2009)\citenamefont {Alayo},
  \citenamefont {Tafur}, \citenamefont {Baggio-Saitovitch}, \citenamefont
  {Pelegrini},\ and\ \citenamefont {Nascimento}}]{ref36}%
  \BibitemOpen
  \bibfield  {author} {\bibinfo {author} {\bibfnamefont {W.}~\bibnamefont
  {Alayo}}, \bibinfo {author} {\bibfnamefont {M.}~\bibnamefont {Tafur}},
  \bibinfo {author} {\bibfnamefont {E.}~\bibnamefont {Baggio-Saitovitch}},
  \bibinfo {author} {\bibfnamefont {F.}~\bibnamefont {Pelegrini}}, \ and\
  \bibinfo {author} {\bibfnamefont {V.~P.}\ \bibnamefont {Nascimento}},\ }\href
  {\doibase 10.1063/1.3122021} {\bibfield  {journal} {\bibinfo  {journal} {J.
  Appl. Phys.}\ }\textbf {\bibinfo {volume} {105}},\ \bibinfo {pages} {093905}
  (\bibinfo {year} {2009})}\BibitemShut {NoStop}%
\bibitem [{\citenamefont {Himi}\ \emph {et~al.}(2001)\citenamefont {Himi},
  \citenamefont {Takanashi}, \citenamefont {Mitani}, \citenamefont {Yamaguchi},
  \citenamefont {Ping}, \citenamefont {Hono},\ and\ \citenamefont
  {Fujimori}}]{ref37}%
  \BibitemOpen
  \bibfield  {author} {\bibinfo {author} {\bibfnamefont {K.}~\bibnamefont
  {Himi}}, \bibinfo {author} {\bibfnamefont {K.}~\bibnamefont {Takanashi}},
  \bibinfo {author} {\bibfnamefont {S.}~\bibnamefont {Mitani}}, \bibinfo
  {author} {\bibfnamefont {M.}~\bibnamefont {Yamaguchi}}, \bibinfo {author}
  {\bibfnamefont {D.~H.}\ \bibnamefont {Ping}}, \bibinfo {author}
  {\bibfnamefont {K.}~\bibnamefont {Hono}}, \ and\ \bibinfo {author}
  {\bibfnamefont {H.}~\bibnamefont {Fujimori}},\ }\href {\doibase
  10.1063/1.1348320} {\bibfield  {journal} {\bibinfo  {journal} {Appl. Phys.
  Lett.}\ }\textbf {\bibinfo {volume} {78}},\ \bibinfo {pages} {1436} (\bibinfo
  {year} {2001})}\BibitemShut {NoStop}%
\bibitem [{\citenamefont {Bloemen}\ \emph {et~al.}(1994)\citenamefont
  {Bloemen}, \citenamefont {van Kesteren}, \citenamefont {Swagten},\ and\
  \citenamefont {de~Jonge}}]{ref29}%
  \BibitemOpen
  \bibfield  {author} {\bibinfo {author} {\bibfnamefont {P.~J.~H.}\
  \bibnamefont {Bloemen}}, \bibinfo {author} {\bibfnamefont {H.~W.}\
  \bibnamefont {van Kesteren}}, \bibinfo {author} {\bibfnamefont {H.~J.~M.}\
  \bibnamefont {Swagten}}, \ and\ \bibinfo {author} {\bibfnamefont {W.~J.~M.}\
  \bibnamefont {de~Jonge}},\ }\href {\doibase 10.1103/PhysRevB.50.13505}
  {\bibfield  {journal} {\bibinfo  {journal} {Phys. Rev. B}\ }\textbf {\bibinfo
  {volume} {50}},\ \bibinfo {pages} {13505} (\bibinfo {year}
  {1994})}\BibitemShut {NoStop}%
\bibitem [{\citenamefont {R$\ddot{o}$${\ss}$ler}\ \emph
  {et~al.}(2006)\citenamefont {R$\ddot{o}$${\ss}$ler}, \citenamefont
  {Bogdanov},\ and\ \citenamefont {Pfleiderer}}]{ref24}%
  \BibitemOpen
  \bibfield  {author} {\bibinfo {author} {\bibfnamefont {U.~K.}\ \bibnamefont
  {R$\ddot{o}$${\ss}$ler}}, \bibinfo {author} {\bibfnamefont {A.~N.}\
  \bibnamefont {Bogdanov}}, \ and\ \bibinfo {author} {\bibfnamefont
  {C.}~\bibnamefont {Pfleiderer}},\ }\href {\doibase 10.1038/nature05056}
  {\bibfield  {journal} {\bibinfo  {journal} {Nature}\ }\textbf {\bibinfo
  {volume} {442}},\ \bibinfo {pages} {797 } (\bibinfo {year}
  {2006})}\BibitemShut {NoStop}%
\bibitem [{\citenamefont {Moriya}(1960)}]{ref25}%
  \BibitemOpen
  \bibfield  {author} {\bibinfo {author} {\bibfnamefont {T.}~\bibnamefont
  {Moriya}},\ }\href {\doibase 10.1103/PhysRev.120.91} {\bibfield  {journal}
  {\bibinfo  {journal} {Phys. Rev.}\ }\textbf {\bibinfo {volume} {120}},\
  \bibinfo {pages} {91} (\bibinfo {year} {1960})}\BibitemShut {NoStop}%
\bibitem [{\citenamefont {Dzyaloshinsky}(1958)}]{ref26}%
  \BibitemOpen
  \bibfield  {author} {\bibinfo {author} {\bibfnamefont {I.}~\bibnamefont
  {Dzyaloshinsky}},\ }\href {\doibase 10.1016/0022-3697(58)90076-3} {\bibfield
  {journal} {\bibinfo  {journal} {J. Phys. Chem. Solids}\ }\textbf {\bibinfo
  {volume} {4}},\ \bibinfo {pages} {241} (\bibinfo {year} {1958})}\BibitemShut
  {NoStop}%
\bibitem [{\citenamefont {Kiselev}\ \emph
  {et~al.}(2011{\natexlab{b}})\citenamefont {Kiselev}, \citenamefont
  {Bogdanov}, \citenamefont {Sch\"afer},\ and\ \citenamefont
  {R\"o\ss{}ler}}]{ref53}%
  \BibitemOpen
  \bibfield  {author} {\bibinfo {author} {\bibfnamefont {N.~S.}\ \bibnamefont
  {Kiselev}}, \bibinfo {author} {\bibfnamefont {A.~N.}\ \bibnamefont
  {Bogdanov}}, \bibinfo {author} {\bibfnamefont {R.}~\bibnamefont {Sch\"afer}},
  \ and\ \bibinfo {author} {\bibfnamefont {U.~K.}\ \bibnamefont
  {R\"o\ss{}ler}},\ }\href {\doibase 10.1103/PhysRevLett.107.179701} {\bibfield
   {journal} {\bibinfo  {journal} {Phys. Rev. Lett.}\ }\textbf {\bibinfo
  {volume} {107}},\ \bibinfo {pages} {179701} (\bibinfo {year}
  {2011}{\natexlab{b}})}\BibitemShut {NoStop}%
\bibitem [{\citenamefont {R$\ddot{o}$$\ss$ler}\ \emph
  {et~al.}(2011)\citenamefont {R$\ddot{o}$$\ss$ler}, \citenamefont {Leonov},\
  and\ \citenamefont {Bogdanov}}]{ref28}%
  \BibitemOpen
  \bibfield  {author} {\bibinfo {author} {\bibfnamefont {U.~K.}\ \bibnamefont
  {R$\ddot{o}$$\ss$ler}}, \bibinfo {author} {\bibfnamefont {A.~A.}\
  \bibnamefont {Leonov}}, \ and\ \bibinfo {author} {\bibfnamefont {A.~N.}\
  \bibnamefont {Bogdanov}},\ }\href@noop {} {\bibfield  {journal} {\bibinfo
  {journal} {J. Phys.: Conf. Ser.}\ }\textbf {\bibinfo {volume} {303}},\
  \bibinfo {pages} {012105} (\bibinfo {year} {2011})}\BibitemShut {NoStop}%
\bibitem [{\citenamefont {Papanicolaou}\ and\ \citenamefont
  {Tomaras}(1991)}]{ref41}%
  \BibitemOpen
  \bibfield  {author} {\bibinfo {author} {\bibfnamefont {N.}~\bibnamefont
  {Papanicolaou}}\ and\ \bibinfo {author} {\bibfnamefont {T.~N.}\ \bibnamefont
  {Tomaras}},\ }\href {\doibase 10.1016/0550-3213(91)90410-Y} {\bibfield
  {journal} {\bibinfo  {journal} {Nucl. Phys. B}\ }\textbf {\bibinfo {volume}
  {360}},\ \bibinfo {pages} {425 } (\bibinfo {year} {1991})}\BibitemShut
  {NoStop}%
\bibitem [{\citenamefont {Tanase}\ \emph {et~al.}(2009)\citenamefont {Tanase},
  \citenamefont {Petford-Long}, \citenamefont {Heinonen}, \citenamefont
  {Buchanan}, \citenamefont {Sort},\ and\ \citenamefont {Nogu\'es}}]{ref33}%
  \BibitemOpen
  \bibfield  {author} {\bibinfo {author} {\bibfnamefont {M.}~\bibnamefont
  {Tanase}}, \bibinfo {author} {\bibfnamefont {A.~K.}\ \bibnamefont
  {Petford-Long}}, \bibinfo {author} {\bibfnamefont {O.}~\bibnamefont
  {Heinonen}}, \bibinfo {author} {\bibfnamefont {K.~S.}\ \bibnamefont
  {Buchanan}}, \bibinfo {author} {\bibfnamefont {J.}~\bibnamefont {Sort}}, \
  and\ \bibinfo {author} {\bibfnamefont {J.}~\bibnamefont {Nogu\'es}},\ }\href
  {\doibase 10.1103/PhysRevB.79.014436} {\bibfield  {journal} {\bibinfo
  {journal} {Phys. Rev. B}\ }\textbf {\bibinfo {volume} {79}},\ \bibinfo
  {pages} {014436} (\bibinfo {year} {2009})}\BibitemShut {NoStop}%
\bibitem [{\citenamefont {Schneider}\ \emph {et~al.}(2000)\citenamefont
  {Schneider}, \citenamefont {Hoffmann},\ and\ \citenamefont {Zweck}}]{ref34}%
  \BibitemOpen
  \bibfield  {author} {\bibinfo {author} {\bibfnamefont {M.}~\bibnamefont
  {Schneider}}, \bibinfo {author} {\bibfnamefont {H.}~\bibnamefont {Hoffmann}},
  \ and\ \bibinfo {author} {\bibfnamefont {J.}~\bibnamefont {Zweck}},\ }\href
  {\doibase 10.1063/1.1320465} {\bibfield  {journal} {\bibinfo  {journal}
  {Appl. Phys. Lett.}\ }\textbf {\bibinfo {volume} {77}},\ \bibinfo {pages}
  {2909} (\bibinfo {year} {2000})}\BibitemShut {NoStop}%
\end{thebibliography}
%

\end{document}